\documentclass{elsart}
\usepackage{epsfig}
\usepackage{amsmath}

 \newtheorem{Theorem}{Theorem}

\usepackage{amssymb}

\begin{document}

\begin{frontmatter}



\title{Stabilization of solitons of the multidimensional nonlinear Schr\"{o}dinger equation: Matter-wave breathers}


\author[UCLM]{Gaspar D. Montesinos},
\author[UCLM]{V\'{\i}ctor M. P\'erez-Garc\'{\i}a},
\author[UGR]{Pedro Torres}

\address[UCLM]{Departamento de Matem\'aticas, Escuela T\'ecnica
Superior de Ingenieros Industriales,
  Universidad de Castilla-La Mancha, 13071 Ciudad Real, Spain}

\address[UGR]{Departamento de Matem\'atica Aplicada, Facultad de Ciencias, \\
  Universidad de Granada, 18071 Granada, Spain}



\begin{abstract}
We demonstrate that stabilization of solitons of the
multidimensional  Schr\"{o}dinger equation with a cubic
nonlinearity may be achieved by a suitable periodic control of the
nonlinear term. The effect of this control is to stabilize the
unstable solitary waves which belong to the frontier between
expanding and collapsing solutions and to provide an oscillating
solitonic structure, some sort of breather-type solution. We
obtain precise conditions on the control parameters to achieve the
stabilization and compare our results with accurate numerical
simulations of the nonlinear Schr\"{o}dinger equation. Because of
the application of these ideas to matter waves these solutions are
some sort of \emph{matter breathers}.
\end{abstract}

\begin{keyword}
Nonlinear Sch\"odinger equations, blow-up, matter waves, solitary
waves
\PACS 03.75. Fi\sep 42.65. Tg\sep 03.40. Kf
\end{keyword}
\end{frontmatter}

\section{Introduction}
\label{I}

The nonlinear Schr\"{o}dinger equation (NLSE) in its many versions
is one of the most important models of mathematical physics, with
applications to different fields such as plasma physics, nonlinear
optics, water waves and biomolecular dynamics, to cite only a few
cases. In many of those examples the equation appears as an
asymptotic limit for a slowly varying dispersive wave envelope
propagating in a nonlinear medium \cite{VVz,Sulem}.

A new burst of interest on problems modelized by nonlinear
Schr\"odinger equations has been triggered by the achievement of
Bose-Einstein condensation using ultracold neutral bosonic gases
\cite{Siempre,Siempre2}.

In particular in Bose-Einstein condensation one of the key
ingredients of the achievement of condensation with alkali gases
 is the trapping of the neutral atoms which
are cooled down below the transition temperatures. Although
different trapping techniques have been used in practice, the most
commonly used are the magnetic traps which are modelized by
external confining forces acting on the system.

There are specific types of Bose-Einstein condensates such as
those made of Lithium \cite{Hulet,Hulet2} in which the
interactions between the atoms are attractive. Mathematically this
implies that for spatial dimensionalities $n=2,3$ collapsing
solutions are possible as it is well known in the framework of the
usual studies of nonlinear Schr\"odinger equations \cite{Sulem}.
Thus, although in principle one could think that nonlinearity and
dispersion could balance mutually and a self-sustained solitonic
structure might exist this is only true for $n=1$. In fact, it was
soon found experimentally \cite{Hulet,Hulet2} and theoretically
supported that such a situation is unstable and leads to collapse
\cite{Sulem,colapsoprimo}. Later studies, taking into account more
elaborate models than the simpler NLS mean field models, led to
the understanding that the occurrence of collapse during the
condensation process would limit the size of an attractive
condensate \cite{Stoof,Hulet3}.

Thus, the only confirmed way to observe solitonic states in
Bose-Einstein condensates (matter-wave solitons) with negative
scattering length involves the elimination of the trap in one
direction as proposed in \cite{Henar,PRAold} and found
experimentally in \cite{nature1,nature2}.

However, the possibility of using Feschbach resonances to control
the scattering length \cite{F1,F2,F3} has provided a way to study
large negative scattering length condensates and collapse
processes in detail \cite{T1,Ueda}. This control allows, with
certain experimental limitations, to modulate appropriately the
nonlinear term.

But this possibility opens new ways for the generation and
observation of different types of matter-wave solitons not yet
fully explored. One of the most striking ones is the possibility
of generating \emph{trapless trapped Bose-Einstein condensate
solitons}. The idea is that (oscillating) bound states might be
obtained by combining cicles of positive and negative scattering
length values so that after an expansion and contraction regime
the condensate would come back to the initial state. In this way
some sort of pulsating trapped condensate, i.e., a
\emph{breather}, would be obtained.

A rigorous study of this intuitive idea is necessary in order to
make precise predictions and find which are the precise parameter
values to be used to obtain such breathers. This is the purpouse
of this paper in a general framework of nonlinear Schr\"{o}dinger
systems. This idea has been explored in two previous papers
\cite{pisaUeda,pisaBoris} but here we improve the understanding of
the phenomenon and correct mistakes contained in the previous
analysis.

Another field of applications of these ideas is nonlinear Optics,
there a full control of the nonlinear term is not possible but,
because of technical limitations, only piecewise constant values
for the nonlinear coefficient can be easily generated
experimentally.

>From the mathematical point of view what we would like is to
stabilize solutions close to the stationary unstable solitons of
the cubic NLSE by choosing appropriate controls of the nonlinear
term. To our knowledge this problem has not been considered
previously in the mathematical literature.

This paper is organized as follows: First, in Sec. \ref{II} we
present the model equations as they appear in one specific
application of the model. Next, in Sec. \ref{III} we reduce the
nonlinear model to a system of ordinary differential equations by
using the so-called moment method. In Sec. \ref{IV} we analyze
two-dimensional systems and compare the analytical predictions of
the moment method with numerical simulations of the full partial
differential equations. We obtain precise conditions for the
existence of breathers and discuss the importance of the initial
data. In Sec. \ref{V} we analyze the three-dimensional case and
explain why  moment (or related) equations fail to describe the
dynamics of the spherically symmetric case. We consider the case
of systems in specific external potentials and give some
stabilization conditions. Finally, in Sec. \ref{VI} we summarize
our conclusions and compare our results with the previous findings
of Refs. \cite{pisaUeda,pisaBoris}.

\section{The model}
\label{II}

In this work we study systems ruled by the NLSE with a cubic
nonlinearity. In the framework of recent problems in Bose-Einstein
condensation this equation appears as the model for the mean field
dynamics of a boson system described by a single wavefunction
$\Psi$ in the zero--temperature limit. The resulting NLSE equation
is called sometimes the Gross--Pitaevskii equation (GPE) which is
\cite{Dalfovo}
\begin{equation}
  \label{eq:gpe}
  i\hbar \frac{\partial \Psi}{\partial \tau} =
  \left[-\frac{\hbar^2}{2m}\triangle + V({\bf r}) +
  \frac{4\pi a \hbar^2}{m}|\Psi|^2\right]\Psi,
\end{equation}
where $\triangle = \sum_{j=1}^3 \partial^2/\partial r_j^2$ is the
three-dimensional Laplacian, $V({\bf r}) =
\frac{1}{2}m\omega^2\sum_{j=1}^3\lambda_j^2r_j^2$ is the external
potential (the so called ``trap") which confines the condensate
and $a$ is the $s$--wave scattering length for the binary
collisions within the condensate.

It is more convenient to work with a new set of adimensional
quantities defined as $x_j = r_j/a_0, t = \tau/T$, $\psi({\bf
x},t) \equiv \Psi({\bf r},\tau)\sqrt{a_0^3/N}$, where
$a_0=\sqrt{\hbar/m\omega}$, $T=1/\omega$ and $N = \int |\Psi|^2
d^3r$ is the number of particles in the condensate. Then Eq.
(\ref{eq:gpe}) reads as the next NLSE
\begin{equation}
  \label{eq:gpe2}
  i \frac{\partial \psi}{\partial t} =
  \left[-\frac{1}{2}\triangle + \frac{1}{2}\sum_{j=1}^3\lambda_j^2x_j^2 + g|\psi|^2\right]\psi,
\end{equation}
where $g = 4\pi N a /a_0$ and $\int |\psi|^2 d^3x = 1$. In this
paper we consider a system evolving without external potential
along one or more dimensions. In Bose-Einstein condensation this
corresponds to a condensate generated in a trap and then the
trapping is eliminated. This leads to different particular cases
of Eq. (\ref{eq:gpe2}). When the potential is removed in all
directions we obtain
\begin{equation}
  \label{eq:gpe3}
  i \frac{\partial \psi}{\partial t} =
  \left[-\frac{1}{2}\triangle + g|\psi|^2\right]\psi.
\end{equation}

 This equation holds for
three-dimensional and quasi two-dimensional systems as will be
discussed in detail in Sec. \ref{V}. Eq. (\ref{eq:gpe3}) has the
typical form of the NLSE with cubic power nonlinearity
\cite{Sulem} and has been extensively studied. In this work we
will analyze the possibilities of controlling the behavior of the
stationary solutions of Eq. (\ref{eq:gpe3}) by letting $g$ be a
time-dependent real function. Similar possibilities for
nonconstant $g$ (in that context z $\leftrightarrow$ t)  but
restricted to $n=2$ arise in the context of coherent light
propagation in nonlinear Kerr media in the paraxial approximation.

\section{Moment method}
\label{III}

In this section we build a theory which will allow us to reduce
the dynamics from the complex model given by Eq. (\ref{eq:gpe3})
to a simpler one whose analysis will provide insight on the basic
features of the phenomenon.

 Let us first consider radially symmetric solutions of Eq.
(\ref{eq:gpe3}) $u=u(r,t)$, $r = (\sum_{j=1}^n x_j)^{1/2}$,
satisfying
\begin{equation}
  \label{modelradial}
  i \frac{\partial u}{\partial t} =
  - \frac{1}{2r^{n-1}} \frac{\partial}{\partial r}\left( r^{n-1}\frac{\partial u}{\partial  r}\right)
  + g(t)|u|^2 u.
\end{equation}
where $n$ is the spatial dimensionality, in our case $n=2,3$. To
get information on the solutions of Eq. (\ref{modelradial}) we
will use the moment method
\cite{Momenta,Momentab,Momentac,momenta2,prl99}. This method
proceeds by analyzing the evolution of several integral quantities
\cite{Momenta,Momentab,Momentac} related with Eq.
(\ref{modelradial}).
\begin{subequations}
  \begin{eqnarray}
   I_1(t) & = & \int |u|^2 d^nx \\
    I_2(t) & = & \int  |u|^2 r^2 d^nx, \\
    I_3(t) & = & i \int  \left(u\frac{\partial u^*}{\partial r} -
      u^*\frac{\partial u}{\partial r}\right)\!r\ d^nx,\\
    I_4(t) & = & \frac{1}{2}\int \left(\left|\nabla u\right|^2
      +\frac{n}{2}g(t) |u|^4 \right) d^n x,\\
      \label{I5} I_5(t) & = & \frac{n}{4}\int |u|^4 d^n x,
  \end{eqnarray}\label{momentos}
\end{subequations}
where $d^nx = r^{n-1} dr d\Omega$. With our scaling for $\psi$,
the first one satisfies $I_1 = 1$, for all $t$. The remaining ones
are related physically to the width, radial momentum and energy of
the wave packet. In what follows we assume that the initial data
are such that all $I_j$ are initially well-defined \cite{SIAMus}.
Some algebra leads to
\begin{subequations}
\begin{eqnarray}
\frac{dI_2}{dt} & = & I_3,  \label{evo_momentos1}\\
\frac{dI_3}{dt} & = & 4 I_4, \\
\frac{dI_4}{dt} & = & g\frac{n-2}{n} \frac{dI_5}{dt} +
\frac{dg}{dt} I_5, \label{evo_momentos4}\\
 \frac{dI_5}{dt} &= & \frac{n\pi 2^{n}}{8}\int \frac{\partial |u|^4}{\partial r}
\frac{\partial\arg u}{\partial r} r^{n-1}dr \simeq -\frac{nI_3
I_5}{2I_2}. \label{evolI5c}
\end{eqnarray}\label{evo_momentos}
\end{subequations}
All the momenta follow closed evolution laws except for $I_5(t)$.
In specific situations the moment equations provide completely
closed equations for the evolution of all the $I_j$
\cite{Momentac,prl99,SIAMus} but this is not our case. Extending
the number of momenta which are included in the calculation does
not help to obtain a closed set of equations.

Thus, to close the system we will take
\begin{equation} \label{cuadratica}
\arg u = I_3r^2/4 I_2,
\end{equation}
as used in \cite{momenta2}. Physically, this corresponds to
approximating the phase of $u$ by the spherical wavefront which
best fits the distribution. A rigorous justification of this
choice is possible \cite{Fibich} for the case $n=2$ and when the
initial data is the so-called Townes soliton to be presented
later.

When solutions with phase given by Eq. (\ref{cuadratica}) are
considered Eqs. (\ref{evo_momentos}) are closed and have several
(positive) invariants under time evolution given by
\begin{subequations}
\begin{eqnarray}
Q_1 & = & 2(I_4-gI_5)I_2-I_3^2/4, \\
Q_2 & = & 2I_2^{n/2}I_5.
\end{eqnarray}\label{Qconservadas}
\end{subequations}
With the help of these quantities, Eqs. (\ref{evo_momentos}) can
be reduced to a single equation for $I_2(t)$, which is
\begin{equation}
\label{nolineali} \frac{d^2 I_2}{dt^2} - \frac{1}{2I_2}
\left(\frac{dI_2}{dt}\right)^2
=2\left(\frac{Q_1}{I_2}+g\frac{Q_2}{I_2^{n/2}}\right).
\end{equation}
If we were able to solve Eq. (\ref{nolineali}) then the use of
Eqs. (\ref{evo_momentos}) would allow us to track the evolution of
the other ones. To simplify Eq. (\ref{nolineali}) we define $X(t)
= \sqrt{I_2(t)}$, which is the wave packet width, and substituting
it into (\ref{nolineali}) we get
\begin{equation}\label{singular}
\ddot{X}=\frac{Q_1}{X^3}+g(t)\frac{Q_2}{X^{n+1}}.
\end{equation}

\section{Two-dimensional systems}
\label{IV}
\subsection{Rigorous analysis of the moment equations}

In physical situations both problems with $n=2$ and $n=3$ arise.
The situation when a Bose-Einstein condensate is tightly confined
along a particular direction leads to a quasi two-dimensional
system such as the ones obtained in Ref. \cite{2DC}. From now on
we consider the two-dimensional model to be valid in that
situation although a more detailed analysis will be made in Sec.
\ref{VI}. Thus, when $n=2$ we obtain
\begin{equation}\label{singular2D}
\ddot{X}=\frac{Q_1 + Q_2 g(t) }{X^3} \equiv \frac{p(t)}{X^3}.
\end{equation}
This equation appears also in relation with the classical motion
of an atom near an infinite straight wire with oscillating charge
(the Paul trap) \cite{P,KL,LZ} and also as an approximate model
for beam propagation in layered Kerr media \cite{Malomed}.

Let us consider the case where $p$ is continuous and $T$-periodic
and look for positive periodic solutions and bound states, that
is, bounded (nonperiodic) positive solutions without collapse. If
$p(t)\geq 0$ all solutions escape to infinity whereas if $p(t)<0$
all solutions collapse. In fact, theorem 2.1 of \cite{KL} implies
that {\em a necessary condition for existence of bound states is}
\begin{equation} \overline p=\frac{1}{T}\int_{0}^{T}p<0.
\end{equation}

Let $p(t)$ be parametrized as $p(t)=\alpha+\beta a(t)$ with
$\overline a=0$. We can fix $a_{M}=\max a(t)=1$ without loss of
generality. A direct consequence of the latter result is that if
$\alpha \geq 0$ bound states cannot exist, therefore $\alpha$ must
be negative. Also, $p(t)$ must change sign, so for existence of
bound states it is necessary that
\begin{equation}\label{necessary} \alpha + |\beta|
>0.
\end{equation}

Let us introduce the positive small parameter
$\epsilon=-2\alpha/3|\beta|$. We will now prove that \emph{there
exists $\epsilon_{0}>0$ (depending only on $a$) such that Eq.
$(\ref{singular2D})$ has a Lyapunov-stable periodic solution if
$0<\epsilon<\epsilon_{0}$}. Moreover, \emph{there is an infinite
number of quasiperiodic solutions and a sequence of subharmonics
with minimal periods tending to $+\infty$}. To prove this
affirmation let us note that when $a(t)=\cos \Omega t$, Theorem 1
in \cite{LZ} asserts that if $\epsilon$ is small enough there is a
periodic solution of twist type \cite{Or}. However, the proof is
still valid for a general continuous and $T$-periodic function
with $\overline a =0$. Moser Twist Theorem \cite{SM} implies that
a solution of twist type is Lyapunov-stable as a consequence of
the existence of invariant curves (quasiperiodic solutions) around
it. Finally, the existence of subharmonics of any order is a
consequence of Poincare-Birkhoff Theorem.

This result has interesting physical consequences. The existence
of periodic solutions is directly related to the existence of
pulsating breathers. The bounded solutions of the theorem would
correspond to solutions with quasiperiodic or chaotic oscillations
in the width. In fact, Mather sets and Smale's horseshoes appear
in the Poincar\'e map. The latter correspond to chaotic
oscillations whereas Mather sets are Cantorian structures that are
more complicated than the quasiperiodic ones.

We have just seen that a necessary condition for the existence of
bounded solutions is that $\alpha+|\beta|>0$ with $\alpha<0$.
Moreover the previous result indicates that the amplitude of the
oscillating term must be of the same order as the amplitude of the
non-oscillating term.

To find particular types of these breathers let us take $g(t)$ as
$g(t)=g_0+g_1 \cos(\Omega t)$ so that $p(t)=\alpha+ \beta
\cos(\Omega t)$ with $\alpha=Q_1+ g_0 Q_2$ and $\beta=g_1 Q_2$. An
useful tool to get periodic solutions is the stability diagram of
Eq. \eqref{singular2D}. We look for the region of parameters for
which there exist stable periodic solutions. To do this we reduce
the number of parameters by defining a new time as
$\tilde{t}=\Omega t$. Then  Eq. \eqref{singular2D} is
\begin{equation}\label{singular2Dn}
X''=\frac{\tilde{\alpha} + \tilde{\beta} \cos(\tilde{t})}{X^3},
\end{equation}
where the prime denotes derivative with respect to $\tilde{t}$,
$\tilde{\alpha}=\alpha/\Omega^2$ and
$\tilde{\beta}=\beta/\Omega^2$. Note that we use the word stable
with two different meanings: the more mathematical one to indicate
Lyapunov-stability and the more physical one to indicate existence
of periodic solutions of the nonlinear Schr\"{o}dinger equation.

\begin{figure}
\begin{center}
\epsfig{file=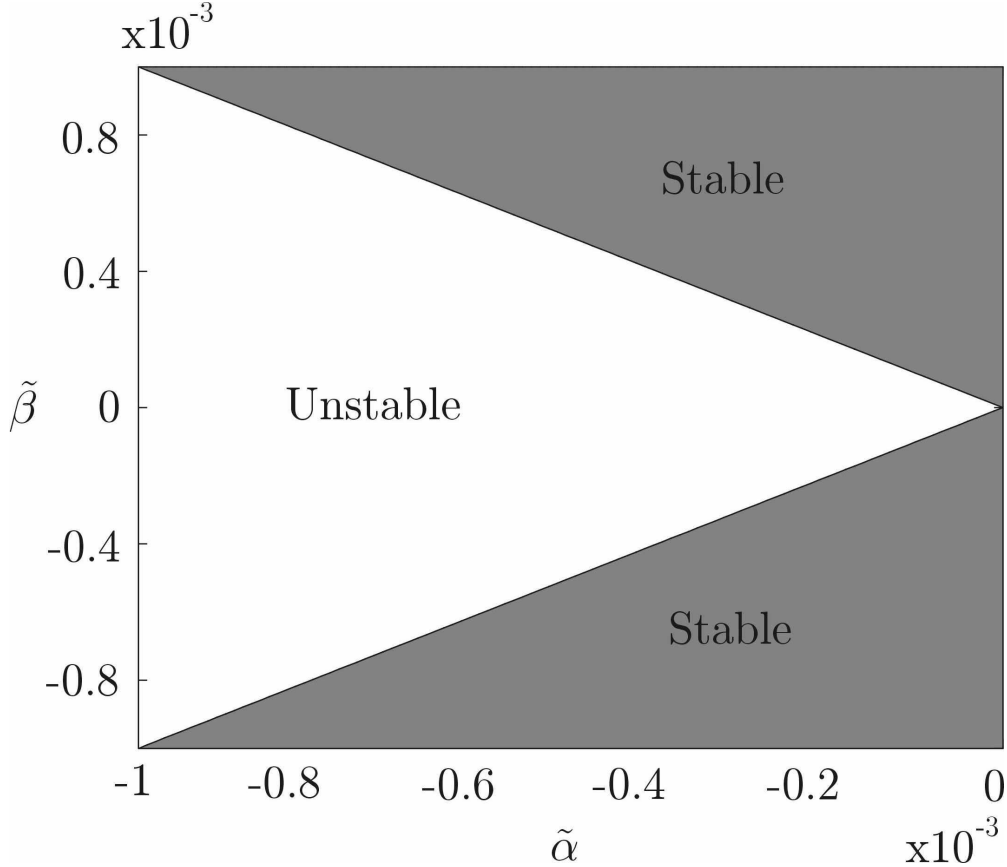,width=0.95\columnwidth}
\end{center}\caption{Stability diagram for Eq. \eqref{singular2Dn} as a function of the
parameters ($\tilde{\alpha}$,$\tilde{\beta}$).\label{diagram}}
\end{figure}

Our objective is to find the region of parameters
($\tilde{\alpha}$,$\tilde{\beta}$) in which there exist initial
conditions $X_0=X(\tilde{t}=0)$, $X_0'=X'(\tilde{t}=0)=0$ for
which Eq. \eqref{singular2Dn} has a stable periodic solution. We
already know that our search can be restricted to the region given
by the conditions $\tilde{\alpha}<0$ and
$\tilde{\alpha}>-|\tilde{\beta}|$. To calculate the stabilization
region we first solve numerically Eq. \eqref{singular2Dn}. Then we
take into account the results that finding a periodic solution of
Eq. \eqref{singular2Dn} is equivalent to finding a solution which
verifies the Neumann conditions $X'(0)=0=X'(2\pi)$. The reason is
that by extending in an even way this solution, we obtain a
4$\pi$-periodic solution that, of course, is bounded and periodic
and this implies that there is a $2\pi$-periodic solution
\cite{Massera}. To obtain solutions that verify the Neumann
conditions, given a pair of parameters
($\tilde{\alpha}$,$\tilde{\beta}$), we look for two initial
conditions $X_{1,0}$ and $X_{2,0}$ for which
$X_1'(2\pi)X_2'(2\pi)<0$. Then, by continuity with respect to the
initial conditions it is known that there exists one initial
condition $X_0$, between $X_{1,0}$ and $X_{2,0}$, for which the
solution of the equation satisfies $X'(2\pi)=0$. We will use two
more observations. First of them is that if $X(\tilde{t})$ is a
solution of the equation with parameters
($\tilde{\alpha}$,$\tilde{\beta}$), then
$Y(\tilde{t})=cX(\tilde{t})$ with $c$ any arbitrary positive
constant is also a solution with parameters ($\tilde{\alpha}
c^4$,$\tilde{\beta} c^4$). Therefore, by moving $c$, all the
points of the line $y=(\tilde{\beta}/\tilde{\alpha})x$ are
obtained and this is nothing but the line which passes through the
points ($\tilde{\alpha}$,$\tilde{\beta}$) and $(0,0)$. So, to scan
the region of parameters we can, for example, to pay attention to
an specific value of $\tilde{\alpha}$ and then change continuosly
the parameter $\tilde{\beta}$. The second observation is that if
$X(\tilde{t})$ is a periodic solution corresponding to a specific
choice of ($\tilde{\alpha}$,$\tilde{\beta}$) then
$Y(\tilde{t})=X(\tilde{t}+\pi)$ is also a solution for parameters
($\tilde{\alpha}$,-$\tilde{\beta}$) since
$\cos(\tilde{t}+\pi)=-\cos(\tilde{t})$. Therefore, the stability
region is symmetric and we can restrict the search to
$\tilde{\beta}>0$. With all these considerations the results
obtained show that the region of parameters for which there exist
initial conditions leading to stable periodic solutions is given
by Eq. (\ref{necessary}) so {\em this condition seems to be not
only necessary but also a sufficient one}. In Fig. \ref{diagram}
this stability region is plotted.

\subsection{Comparison with full numerical simulations of the NLSE}

In this subsection we will compare the predictions based on the
ordinary differential equation \eqref{singular2D} with the
simulations of the full nonlinear Schr\"{o}dinger equation
\eqref{eq:gpe3}.

First of all we have to choose appropriate initial data
$u_0=u(r,t=0)$ in order to solve numerically the NLSE. Since we
are going to try to stabilize the solutions it is reasonable to
start trying to stabilize the so-called stationary solutions. So,
let us consider stationary solutions of Eq. \eqref{eq:gpe3} when
$g=g_s$ is constant, given by $u(\textbf {x},t)=e^{i\mu t}
\Phi_{\mu}(\textbf{x})$. Here $\Phi_{\mu} (\textbf{x})$ verifies
\begin{equation}\label{soliton}
\triangle \Phi_{\mu} -2\mu \Phi_{\mu}- 2 g_s |\Phi_{\mu}|^2
\Phi_{\mu}=0.
\end{equation}
As it is precisely stated in \cite{Sulem}, when $g_s$ is negative,
for each positive $\mu$ there exists only one solution of Eq.
\eqref{eq:gpe} which is real, positive and radially symmetric and
for which $(\int |\Phi_{\mu}|^2 d^n x)^{1/2}$ has the minimum
value between all of the possible solutions of Eq.
\eqref{soliton}. Moreover, the positivity of $\mu$ ensures that
this solution decays exponentially at infinity. This solution is
called the ground state and, in two dimensions, is known as the
Townes soliton. We will denote this solution as $R_{\mu}(r)$ and
satisfies the equation
\begin{equation}\label{solitonR}
\triangle R_{\mu} -2\mu R_{\mu}- 2 g_s R_{\mu}^3=0
\end{equation}
and the boundary conditions
\begin{equation}\label{boundary}
\lim_{r\rightarrow{\infty}} R_{\mu}(r)=0,\ R_{\mu}'(0)=0.
\end{equation}
Fixed a value of $\mu$, the norm and width of $R_{\mu}$ are given
by $\eta_{\mu}=(\int |R_{\mu}|^2 d^2 x)^{1/2}$ and $X_{\mu}=(\int
|R_{\mu}|^2 r^2 d^2 x)^{1/2}$ respectively. Applying scaling
trasformations to $R_{\mu}(r)$ it is possible to build stationary
solutions having the same shape and norm as the Townes soliton but
different widths
\begin{equation}
R_{\mu} (r)=\mu^{1/2} R_1 (\mu^{1/2} r),\ X_{\mu}=X_1/\mu^{1/2}.
\end{equation}
The equation which verifies the normalized soliton $R_{\mu, N}
(r)=R_{\mu} (r)/\eta_{\mu}$ is
\begin{equation}\label{solitonN}
\triangle R_{\mu, N} -2\mu R_{\mu, N}- 2 g_s \eta_{\mu}^2 R_{\mu,
N}^3=0.
\end{equation}
We have taken as initial condition $u(r,t=0)=R_{\mu,N}(r)$ for
some $\mu$ (therefore the initial width is fixed) and $g_s$
values. To obtain the shape $R_{\mu}(r)$ we have used a shooting
method \cite{Jimenez} to solve Eq. \eqref{solitonR}. The idea is
to rewrite Eq. \eqref{solitonR} as a dynamical system and impose
that the solution of such a system also verifies the boundary
conditions \eqref{boundary}. In order to avoid the singularity
which appears at the point $r=0$ we have solved the
indetermination by means of a series expansion of $R_{\mu}$ around
$r=0$.

>From the theory of nonlinear Schr\"odinger equations it is known
that the Townes soliton is unstable, i.e. small perturbations of
this solution lead to either expansion of the initial data or
blow-up in finite time. Thus, following the analogy with the
stabilization of unstable fixed points in finite dimensional
dynamical systems by fast variation of a parameter, we will try to
stabilize this unstable but anyway stationary solution of Eq.
(\ref{eq:gpe3}).

Using the predictions of Eq. \eqref{singular2D}, in order to get
stable periodic solutions, we have to choose $g_0$ such that
$\alpha=Q_1+g_0 Q_2$ is negative, that is, the following relation
must hold
\begin{equation}\label{g0colapso}
g_0 < -\frac{\int |\nabla u_0|^2 d^2 x}{\int |u_0|^4 d^2 x}.
\end{equation}
When only such a $g_0$-term is present the solutions collapse as
predicted by Eq. \eqref{singular2D}. So, the $g_1$-term is
necessary in order to arrest the collapse and achieve the
stabilization of the soliton given by $u_0$.

Thus, we have first taken as initial data the Townes soliton
obtained for $g_s=-0.5$ and $\mu=0.5$ whose width is $X_0=1.09$
and its norm satisfies $g_s \eta_{\mu=0.5}^2=-5.85$. This value is
precisely the critical nonlinearity for collapse below which the
collapse occurs. Let us note that only for the Townes soliton it
is satisfied  that the collapse threshold in Eq. \eqref{eq:gpe3},
when $g$ is a constant, is just the value of the nonlinearity in
Eq. \eqref{solitonN} \cite{Sulem,Fibich}.

\begin{figure}
\begin{center}
\epsfig{file=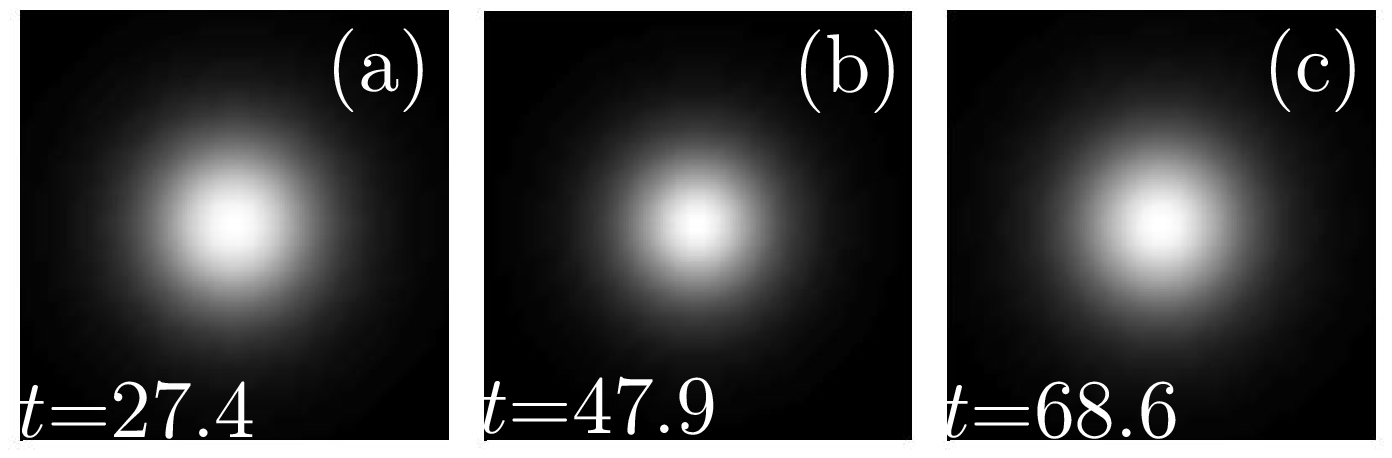,width=0.95\columnwidth}
\epsfig{file=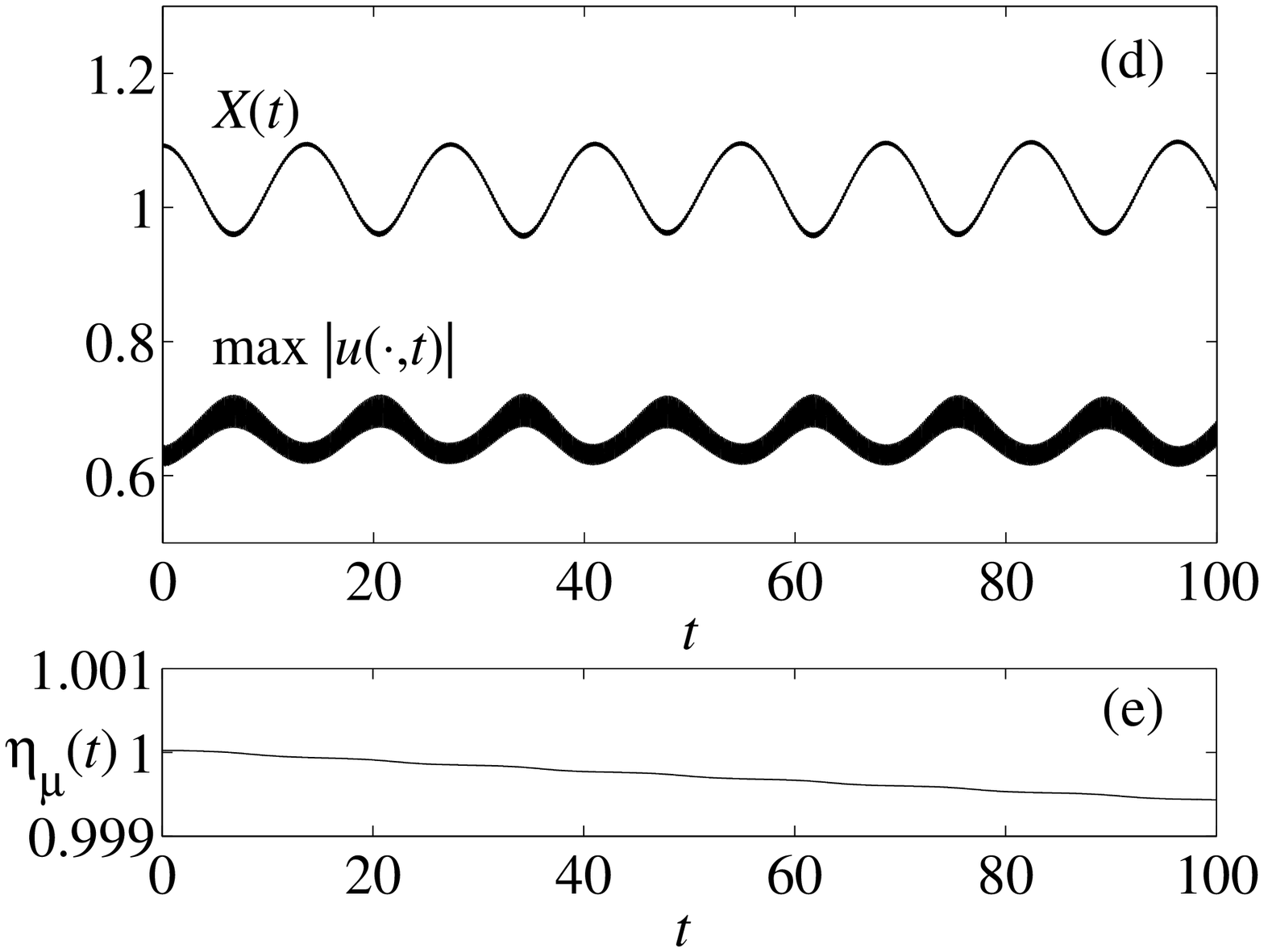,width=0.95\columnwidth}
\end{center}\caption{Results of numerical simulations of Eq. \eqref{eq:gpe3}
showing stabilization of the initial data $u({\bf
x},0)=R_{\mu,N}({\bf x})$ with $\mu=0.5$ and $g_s=-0.5$
($X_0=1.09$) for parameter values $g_0=-2\pi$, $g_1=8\pi$,
$\Omega=40$. (a-c) Plots of $|u({\bf x},t)|^2$ on the spatial
region $[-2,2]\times [-2,2]$ corresponding to times of maximum
(a,c) and minimum (b) width of the solution. (d) Evolution of the
width $X(t)$ and of the amplitude $\max |u(\cdot,t)|$. (e)
Evolution of the norm $\eta_{\mu}(t)$.\label{soliton2d}}
\end{figure}

In all the simulations of Eq. \eqref{eq:gpe3} we have made there
appear two different types of behaviors. First, there is a
small-amplitude fast oscillation due to the term $\cos(\Omega t)$
which is present in both the full model Eq. (\ref{eq:gpe3}) and
the reduced ODE system. Secondly, there is a ``slow" dynamics due
to the internal dynamics of the system which sometimes does not
appear in the ODE model. When the results from the ODE
\eqref{singular2D} and the full PDE are compared qualitatively
similar dynamics are observed in many situations. For instance, in
Fig. \ref{soliton2d} we plot the results of the simulation for
parameter values $g_0=-2\pi$, $g_1=8\pi$ and $\Omega=40$. To solve
numerically Eq. \eqref{eq:gpe3} we have used a pseudo-spectral
method combined with a second order split-step method to advance
in time \cite{Taha,Blanes,Xiaoyan}. We have also compared this
simulation with the one obtained with a fourth order split-step
method and found both results in full agreement.

It is important to indicate that it is absolutely necessary to
incorporate an absorbing potential at the frontier of the
simulation region to avoid possible interferences between the
fraction of the wave packet moving outside (i.e. a possible
non-trapped fraction of initial data) and the fraction of the
initial data which remains trapped. Typical grid sizes were
$256\times 256$ and the time step was $dt=0.001$. All results were
tested on different grid sizes and changing time steps.

Clearly, the system is trapped with a fast modulation in the
nonlinearity ($g_1$-term) and the collapse process is inhibited.
Also, there is a fast oscillation with the same frequency $\Omega$
as the one of the oscillating term,  (beyond the resolution of the
plot), and a slow oscillation due to the proper dynamics of the
system.

In Fig. \ref{soliton2d}(e) it is observed that the norm of the
solution decreases in time. This is caused by the action of the
absorbing potential on the outgoing way.

\begin{figure}
\begin{center}
\epsfig{file=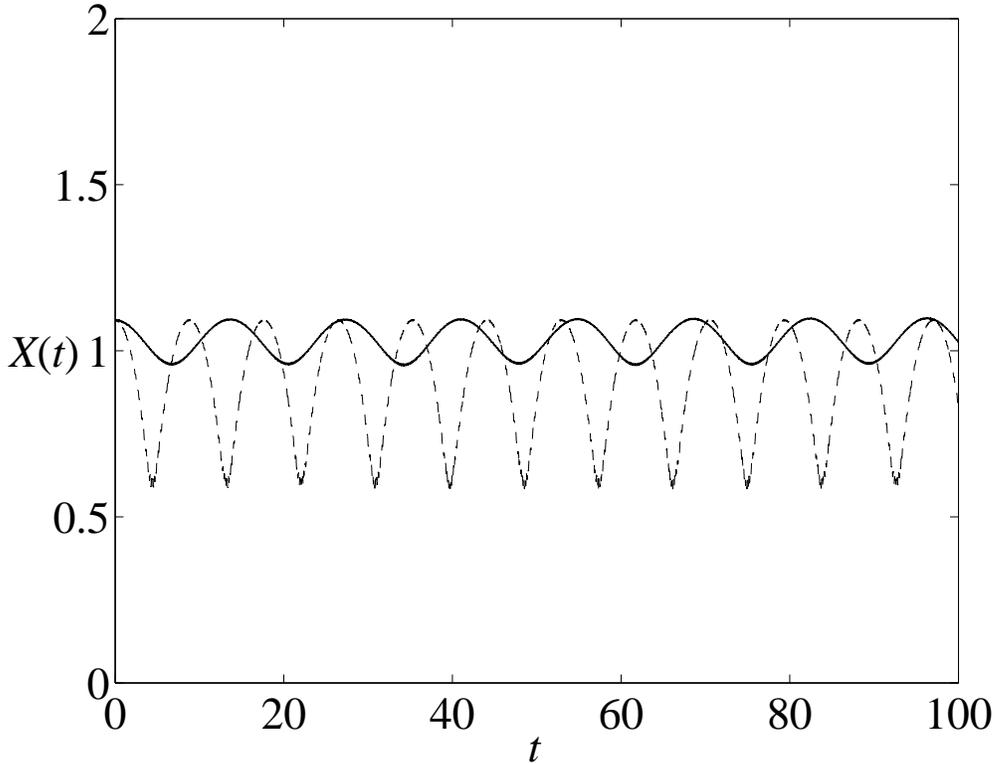,width=0.95\columnwidth}
\end{center}\caption{Results of numerical simulations of Eq. \eqref{eq:gpe3}
 and Eq. \eqref{singular2D} showing stabilization
of the initial data $u({\bf x},0)=R_{\mu,N}({\bf x})$ with
$\mu=0.5$ ($X_0=1.09$) and $g_s=-0.5$ for parameter values
$g_0=-2\pi$, $g_1=8\pi$, $\Omega=40$. The evolution of the width
$X(t)$ according to Eq. \eqref{eq:gpe3} (solid) and to Eq.
\eqref{singular2D} (dashed) is shown.\label{comparacion2d}}
\end{figure}

The next step in our study is to compare this stabilization in the
NLSE with the differential equation for the width. In Fig.
\ref{comparacion2d} we see that the quantitative predictions of
Eq. \eqref{singular2D} are not very precise as what concerns the
amplitude and frequency of the slow oscillation. In any case, at
least the simple ODE model predicts correctly the trapping of the
solution for these parameters. These results imply that the
predictions of the ODE system must be taken only as qualitative
indications of the possible dynamics.

\begin{figure}
\begin{center}
\epsfig{file=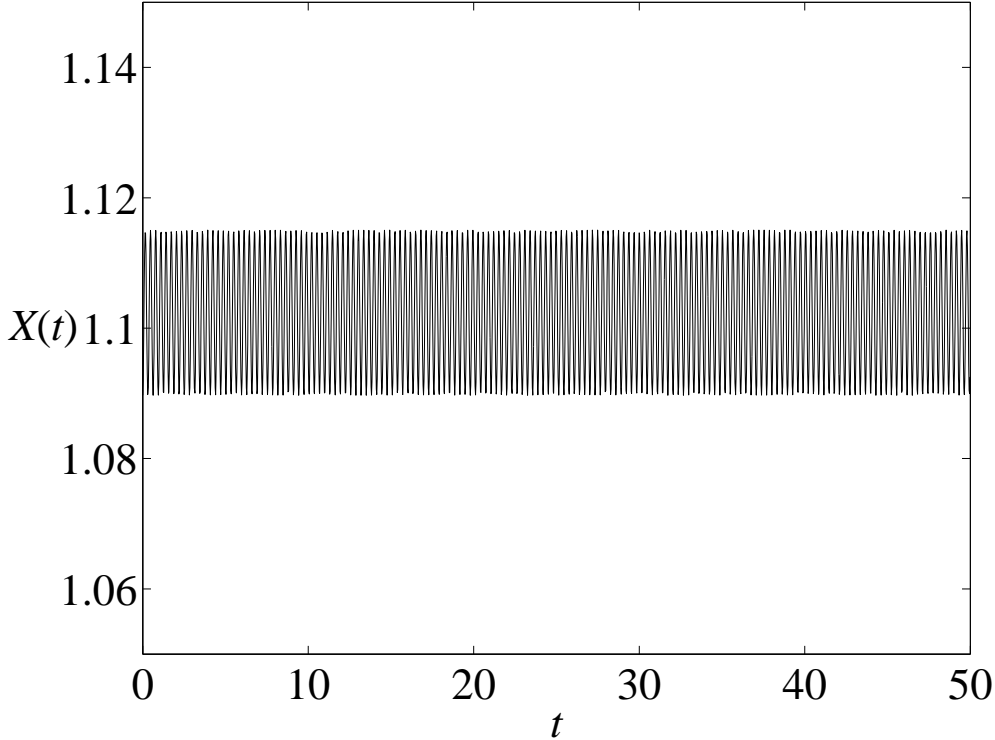,width=0.95\columnwidth}
\end{center}\caption{Results of numerical simulations of Eq. \eqref{singular2D}
showing stabilization of the initial data $u({\bf
x},0)=R_{\mu,N}({\bf x})$ with $\mu=0.5$ and $g_s=-0.5$
($X_0=1.09$) for parameter values $g_0=-6.4$, $g_1=33.4$,
$\Omega=20$. The evolution of the width $X(t)$ is plotted.
\label{fastsoliton}}
\end{figure}

Also, there exist parameter ($g_0$, $g_1$ and $\Omega$) choices
which lead to stabilization in Eq. \eqref{singular2D} but not in
Eq. \eqref{eq:gpe3}. For example, according to Eq.
\eqref{singular2D} and the diagram of Fig. \ref{diagram} the
parameters $g_0=-6.4$, $g_1=33.4$ and $\Omega=20$ stabilize the
initial soliton as shown in Fig. \ref{fastsoliton}. However, this
stabilization does not happen when we solve numerically Eq.
\eqref{eq:gpe3}, instead, after a few oscillations their dynamics
are drastically different. 

\subsection{Stabilization of other initial data}

 Although a Townes soliton is a natural object to stabilize in
 the framework of Eq. (\ref{eq:gpe3}) we may wonder (i) if the
 procedure described above does stabilize other type of initial
 data and (ii) if the stabilization is appropriately described by Eqs.
 \eqref{singular2D}.

 From the point of view of applications of Bose-Einstein
 condensation the possibility of stabilizing other initial data is
 important since it is not clear how a Townes soliton could be
 generated in real experiments. On the other hand, other initial
 data  such as Thomas-Fermi type solutions
 or gaussians are much more natural and easier to obtain. This,
 together with the fact
 that the usual time dependent variational method \cite{PRAold} is
 usually developed for gaussian profile functions has lead to some
 interest on the possibility of trapping gaussian initial data
 (which is the case studied in Refs. \cite{pisaUeda,pisaBoris}).

Therefore, let us take $u_0$ as
\begin{equation}
u_0 (r)=\frac{e^{-r^2/2 X_0^2}}{\sqrt{\pi}X_0},
\end{equation}
for which
\begin{subequations}
\begin{eqnarray}
\eta & = & \left(\int |u_0|^2 d^2x\right)^{1/2}=1,\\
 X_0 & = & \left(\int
|u_0|^2 r^2 d^2x\right)^{1/2}.
\end{eqnarray}
\end{subequations}
In this case we get for the invariants the values $Q_1=1$ and
$Q_2=1/2\pi$ independently of the width $X_0$. So, the equation
for the width is
\begin{equation}\label{singulargauss}
\ddot{X}=\frac{1 +g(t)/2\pi}{X^3} \equiv \frac{p(t)}{X^3}.
\end{equation}
which corresponds to the evolution equation of Refs.
\cite{pisaUeda,pisaBoris}. When $g$ is a constant this equation
predicts a threshold value for collapse of $g=-2\pi$ which is,
indeed, below the real value which corresponds to the Townes
soliton ($g_s\eta_{\mu}^2=-5.85$). This means that when a Gaussian
function is taken as initial data, Eq. \eqref{singulargauss} does
not describe accurately the region of trapping, because for
$-2\pi<g_0<-5.85$ this equation predicts expansion of the initial
data while the real dynamics of the partial differential equation
is collapsing.

\begin{figure}
\begin{center}
\epsfig{file=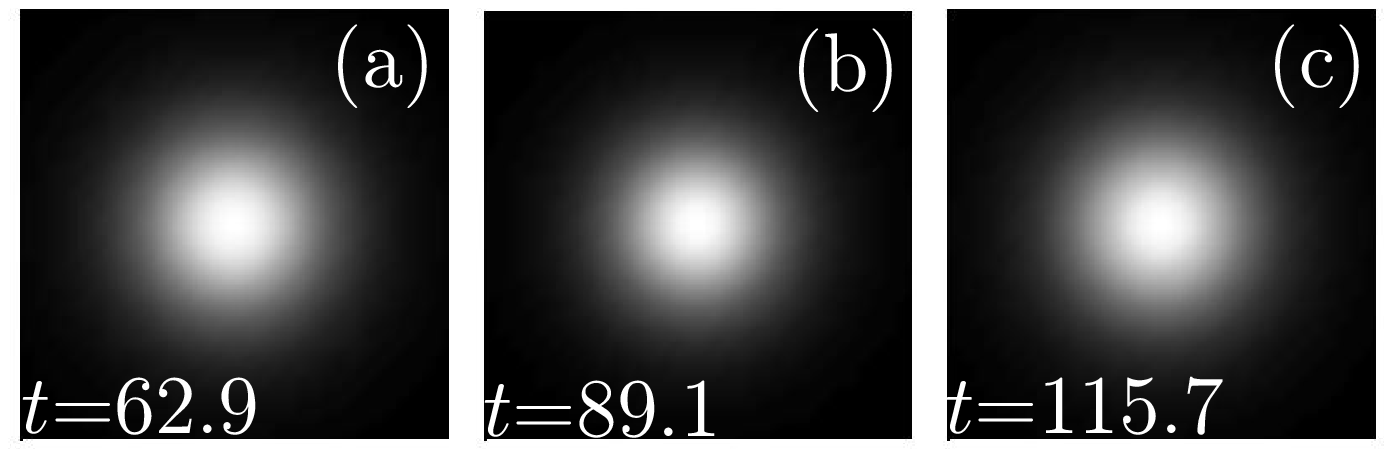,width=0.95\columnwidth}
\epsfig{file=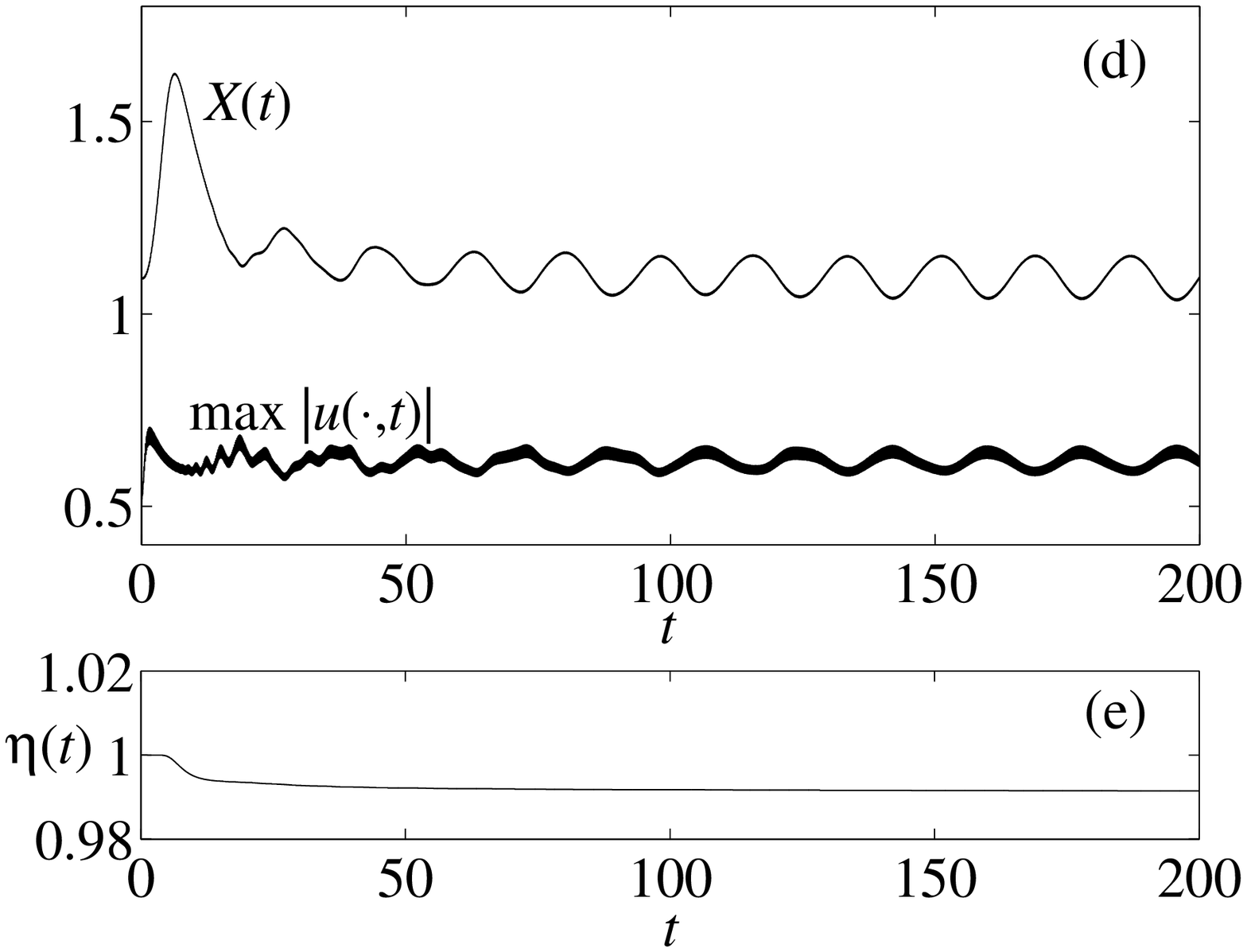,width=0.95\columnwidth}
\end{center}\caption{Results of numerical simulations of Eq. \eqref{eq:gpe3}
showing stabilization of Gaussian initial data $u({\bf
x},0)=(1/\sqrt{\pi}X_0) e^{-r^2/2 X_0^2}$ with $X_0=1.09$ for
parameter values $g_0=-2\pi$, $g_1=8\pi$, $\Omega=40$. (a-c) Plots
of $|u({\bf x},t)|^2$ on the spatial region $[-2,2]\times [-2,2]$
corresponding to times of maximum (a,c) and minimum (b) width of
the solution. (d) Evolution of the width $X(t)$ and of the
amplitude $\max |u(\cdot,t)|$. (e) Evolution of the norm
$\eta(t)$.\label{gauss2d}}
\end{figure}

In Fig. \ref{gauss2d} the evolution of a Gaussian with $X_0=1.09$
and parameter values $g_0=-2\pi$, $g_1=8\pi$ and $\Omega=40$ is
shown. Notice that these are the same width and parameters which
allow trapping of a Townes soliton.

We can see that with these parameters the stabilization is
achieved although Eq. \eqref{singulargauss} does not predict
trapping (in fact for these parameters $\bar{p} = 0$). What is the
difference with the stabilization of the Townes soliton shown in
Fig. \ref{soliton2d}?. A first important observation is that in
the case of  Gaussian initial data a readjustment is produced soon
by ejecting a significant part of the wave packet far from the
trapping region. This corresponds to the first 10 time units where
the width significantly increases due to the contribution of the
outgoing wave. When this wave hits the absorbing region it is
dissipated leading to the step in the norm evolution shown in Fig.
\ref{gauss2d}(e). What remains trapped is indeed a Townes soliton.

Thus, the use of initial data different from a Townes soliton
leads to a splitting of the solution into the soliton itself,
which is the structure which can be stabilized, plus a certain
amount of radiation which goes far from the region of interest.
Because of this fact one must be careful to eliminate the outgoint
part of the radiation in the numerical simulations. We have
verified that when the absorbing region is absent the numerical
simulations are misleading and the trapping effects are
drastically altered (in fact, when zero boundary conditions at a
given distance are used, there appear reflections in the boundary
and essential changes on the results take place which would lead
to spurious destabilization of the system after very short
trapping times).

 It is known that Eq. (\ref{eq:gpe3}) with a cubic
nonlinearity and two spatial dimensions corresponds to the
so-called critical case \cite{Sulem,Fibich} in which diffraction
and self-focusing are nearly balanced and collapse is extremely
sensitive to perturbations and to changes in the initial
conditions. For this reason although Gaussians look roughly like
the Townes soliton they are not able to capture the delicate
balance between diffraction and nonlinear focusing present in this
case. When a Townes soliton is taken as a initial condition in Eq.
\eqref{eq:gpe3} there is only a very small outgoing component
which spreads out (due to numerical errors) and the system
responds to the parametric perturbation as a whole. This fact
implies that the moment equations may describe reasonably well the
dynamics of the system only near the Townes soliton, since they
only deal the global dynamics of the system. However, when the
initial condition is a Gaussian, the component which spreads out
cannot be captured neither by a moment-type formalism nor by the
usual variational methods.

\section{Three-dimensional systems}
\label{V}
\subsection{Analysis of the moment equations}

Let us now consider the case $n=3$, then Eq. (\ref{singular})
reads
\begin{equation}\label{singular3D}
\ddot{X}=\frac{Q_{1}}{X^3}+g(t)\frac{Q_{2}}{X^4}.
\end{equation}
Let us first assume that $g$ is continuous and $T$-periodic, then
{\em if $\overline g=\frac{1}{T}\int_{0}^{T}g\geq 0$ bound states
cannot exist}. The reason is that, by Massera theorem, a bound
state would imply the existence of a periodic solution. But if
there is a periodic solution $X$, multiplying $(\ref{singular3D})$
by $X^4$ and integrating over a period we get $
0>-4\int_{0}^{T}X^3\dot{X}^2-Q_{1}\int_{0}^{T}X=\overline g Q_{2}
T\geq 0$, which is a contradiction. Therefore, {\em a necessary
condition for the existence of bounded solutions is}
\begin{equation}
\overline g=\frac{1}{T}\int_{0}^{T}g < 0.
\end{equation}

Note that this result is the opposite to the one inferred in Ref.
\cite{pisaBoris} from direct numerical simulations an equation
similar to (but restricted to the class of gaussian initial data)
 that of Eq. (\ref{singular3D}).

Our goal is to find stable periodic solutions of Eq.
(\ref{singular3D}) in three dimensions. A first useful observation
is that an arbitrary positive $T$-periodic function $X(t)$ is a
solution of $(\ref{singular3D})$ if
$g(t)=Q_{2}^{-1}(X^4\ddot{X}-Q_{1}X)$. In the following result, we
use the definition of positive part of a function as
$f(t)^+=\max\{0,f(t)\}$.

\begin{Theorem} Let $X(t)$ be a $T$-periodic positive function such
that

(i) $\displaystyle \int_{0}^{T}
 4\frac{\ddot{X}}{X}-\frac{Q_{1}}{X^4}>0$

(ii) $\displaystyle T\int_{0}^{T}\frac{\ddot{X}^+}{X}<1$

Then, there exists $g(t)$ $T$-periodic function such that $X(t)$
is a $T$-periodic linear stable solution of $(\ref{singular3D})$.
\end{Theorem}

{\noindent{\bf Proof. }} If we choose
$g(t)=Q_{2}^{-1}(X^4\ddot{X}-Q_{1}X)$, then we know that $X(t)$ is
a $T$-periodic solution of $(\ref{singular3D})$. The linearized
equation is
$$
\ddot{Y}+\left(4\frac{\ddot{X}}{X}-\frac{Q_{1}}{X^4}\right)Y=0.
$$
Now, hypotheses $(i)-(ii)$ correspond to the assumptions of the
classical Lyapunov's criterion for linear stability (see for
instance Theorem 1 in \cite{Zhang}).

In many cases, hypotheses $(i)-(ii)$ can be verified numerically.
For instance taking $g(t)=-Q_{2}^{-1}(15+\sin t)^4 \sin t
-Q_{2}^{-1}Q_{1}(15+\sin t)$, then equation $(\ref{singular3D})$
has the solution $X(t)=15+\sin t$ which is linearly stable.

A different way to build solutions to Eqs. (\ref{singular3D}) is
to choose $g(t) = G(t) X(t)$ then Eq. (\ref{singular3D}) becomes
equivalent to Eq. (\ref{singular2D}) and thus all the results of
existence of breathers in two dimensions can be applied.

In conclusion, from the analysis of Eq. \eqref{singular3D} we get
that trapping could be possible in a three-dimensional scenario
provided the dynamics of the system is well described by the
moment equations (\ref{singular3D}).

\subsection{Failure of moment equations for symmetric 3D systems}

To check the validity of the predictions made on the basis of Eq.
(\ref{singular3D}) we have compared its predictions with the
behavior of the solution as obtained from direct numerical
simulations of Eq. \eqref{eq:gpe3}. We have used as initial data
solutions of Eq. \eqref{solitonN} for some $\mu$ and $g_s$ values
obtained with a shooting method as we described above for the
two-dimensional case. When $n=3$, the relations between soliton
solutions of different widths are
\begin{equation}
R_{\mu} (r)=\mu^{1/2} R_1 (\mu^{1/2} r),\ X_{\mu}=X_1/\mu^{3/4}.
\end{equation}
The norm is not conserved by the previous change and verifies the
relation $\eta_{\mu}=\eta_{1}/\mu^{1/4}$.

In all the simulations we have taken $g(t)=g_0+g_1 \cos(\Omega t)$
with parameters $g_0, g_1$ and $\Omega$ chosen to give stable
behavior on the basis of  Eq. \eqref{singular3D},

Our numerical results show that the stabilization predicted by Eq.
(\ref{singular3D}) does not occur for solutions of Eq.
(\ref{eq:gpe3}). Although it is possible by a suitable choice of
 $g_0$, $g_1$ and $\Omega$ to change the behavior from collapsing to expanding or
 viceversa from the behavior which corresponds to the $g_1= 0$ case, we were not
 able to find stable breather solutions for any choice of the parameters. In the
region in which the solutions change from collapsing to expanding
it is possible to
  fine-tune the parameters to get a solution which holds for some
  time, but this structure is unstable and finally
the solution loses its profile.

Thus, the moment equations fail to describe the dynamics of the
system in three dimensions. The reason is simple: when the
nonlinear term in Eq. (\ref{eq:gpe3}) is cubic and $n=3$ we are in
the so-called supercritical case in which collapse occurs by
formation of a localized spike on a finite-amplitude background.
This behavior cannot be captured neither with the quadratic phase
approximation [Eq. (\ref{cuadratica})] nor with any type of
variational ansatz.

\subsection{Three dimensional systems with confinement along one direction}

Although stabilization of spherical structures seems not possible
in three-dimensional scenarios, we could think about the situation
in which the three-dimensional system has cylindrical symmetry. It
would seem plausible that if the system is coin-shaped
stabilization might take place because, in fact, this system is
close to a quasi two-dimensional one and, therefore, the results
in two dimensions could be applied. In fact, a numerical
simulation reported in Ref. \cite{pisaUeda} supports this
conjecture. Here we will try to make a more systematic analysis of
the phenomenon.

To get a qualitative understanding of the phenomenology for this
situation let us consider Eq. \eqref{eq:gpe2}. We will take
gaussian functions as initial data and obtain the evolution
equations for the parameters by the use of the collective
coordinates method \cite{PRAold} to find the evolution of the
widths in each spatial direction. The idea is to restate the
problem of solving Eq. \eqref{eq:gpe2} as a variational problem,
corresponding to a stationary point of the action related to the
Lagrangian density $\mathcal{L}$. So, the problem is transformed
into the problem of finding $\psi(\textbf{x},t)$ such that the
action
\begin{equation}
S=\int \mathcal{L}\ d^nx dt
\end{equation}
is extreme. This problem is as complicated as solving the original
NLSE. The idea of the method is to restrict the analysis to a set
of trial functions. One possible choice is to take a Gaussian
ansatz of the form
\begin{equation}\label{ansatz}
\psi({\bf x},t)=\frac{1}{\pi^{n/4} \prod_{j=1}^n w_j(t)^{1/2}}
\prod_{j=1}^n e^{-x_j^2/2 w_j^2+i x_j \alpha_j(t)+i x_j^2
\beta_j(t)},
\end{equation}
which verifies the following relations
\begin{subequations}
\begin{eqnarray}
\eta(t)& = & \left(\int |u|^2 d^n x\right)^{1/2}=1, \\
w_j (t) & = & 2 \int |u|^2 x_j^2 d^n x.
\end{eqnarray}
\end{subequations}
The variational method leads to the following evolution equations
for the widths
\begin{equation}
\ddot{w_j}+\lambda_j^2(t)
w_j=\frac{1}{w_j^3}+\frac{g(t)/(2\pi)^{n/2}}{w_j}\left(\prod_{k=1}^n
\frac{1}{w_k}\right).
\end{equation}
In the three-dimensional case the equations are
\begin{subequations}
\begin{eqnarray}
\ddot{w_1}+\lambda_1^2(t) w_1 & = &
\frac{1}{w_1^3}+\frac{g(t)/(2\pi)^{3/2}}{w_1^2 w_2 w_3},\\
\ddot{w_2}+\lambda_2^2(t) w_2 & = &
\frac{1}{w_2^3}+\frac{g(t)/(2\pi)^{3/2}}{w_1 w_2^2 w_3},\\
\ddot{w_3}+\lambda_3^2(t) w_3 & = &
\frac{1}{w_3^3}+\frac{g(t)/(2\pi)^{3/2}}{w_1 w_2 w_3^2}.
\end{eqnarray}
\end{subequations}
If we consider cylindrically symmetry solutions we have
$w_1=w_2=w$, this width $w$ being equivalent to the width $X$
arising in the radially symmetric two-dimensional case.
\begin{equation}
X=\left(\int |u|^2 (x^2+y^2)
d^3x\right)^{1/2}=\left(\frac{w_1^2+w_2^2}{2}\right)^{1/2}=w.
\end{equation}
Therefore, if we consider the case where the potential is present
only along the $z$ direction and we suppose cylindrical symmetry
the equations for the widths are
\begin{subequations}\label{eq:3dwidths}
\begin{eqnarray}
\ddot{w} & = & \frac{1}{w^3}+\frac{g(t)/(2\pi)^{3/2}}{w^3 w_3},\label{eq:wcil3d}\\
\ddot{w_3}+\lambda_3^2(t) w_3 & = &
\frac{1}{w_3^3}+\frac{g(t)/(2\pi)^{3/2}}{w^2 w_3^2}.\label{eq:w3}
\end{eqnarray}
\end{subequations}
Our goal is to stabilize some solutions of this model, i.e. to get
periodic solutions for $(w(t),w_3(t))$. In particular, if we
impose that
\begin{equation}\label{eq:g3d}
\frac{g(t)}{(2\pi)^{3/2} w_3(t)}=\frac{g_{2D}(t)}{2\pi},
\end{equation} where
$g_{2D}(t)$ is an specific modulation of the nonlinear term
allowing trapping in two dimensions, then Eq. \eqref{eq:wcil3d} is
the same equation that in the radially symmetric two-dimensional
case (Eq. \eqref{singulargauss}) and, therefore, by taking the
three-dimensional modulation of the form of Eq. \eqref{eq:g3d} we
will have stabilization of the width $w$. Now, Eq. \eqref{eq:g3d}
can be written as
\begin{equation}\label{eq:g3dg2d}
g(t)=(2\pi)^{1/2} g_{2D}(t) w_3(t),
\end{equation}
and the problem arises when we realize that $w_3(t)$ is not known,
but it evolves according to Eq. \eqref{eq:w3}, so Eq.
\eqref{eq:g3dg2d} is useless in practical cases. Nevertheless, we
could think about the possibility of stabilizing $w_3$ around the
equilibrium point of Eq. \eqref{eq:w3} if the nonlinear term were
absent. This equilibrium point is $w_{3,0}=\lambda_3^{-1/2}$. To
ensure that the value of $w_3$ is as similar to the equilibrium
value as possible, we have to impose that
\begin{equation}
\left|\frac{g(t)/(2\pi)^{3/2}}{w(t)^2 w_{3,0}^2}\right| \ll
\frac{1}{w_{3,0}^3},
\end{equation}
for all $t$, and this implies that
\begin{equation}\label{eq:lambda}
\lambda_3 \gg \max_t\left(\left|\frac{g_{2D}(t)}{2\pi
w^2(t)}\right|\right).
\end{equation}
If Eq. \eqref{eq:lambda} were satisfied we could take
$w_3(t)\simeq w_{3,0}=\lambda_3^{-1/2}$ in Eq. \eqref{eq:g3dg2d}
and from this equation we get a prediction for the values of the
modulation we should take to get stabilization. As we know the
values of $g_{2D}$ and $w$ to stabilize in two dimensions, Eq.
\eqref{eq:lambda} is the condition to get stabilization in three
dimensions (with the potential along $z$). As we said before, this
condition means that the more coin-shaped the system is, the
better the stabilization will be, leading to a quasi
two-dimensional system. Again, the variational method predicts
that stabilization can occur only if $g_0<(2\pi)^{1/2}(-2\pi)$ but
we will see later that it is also possible for other $g_0$ values.

\begin{figure}
\begin{center}
\epsfig{file=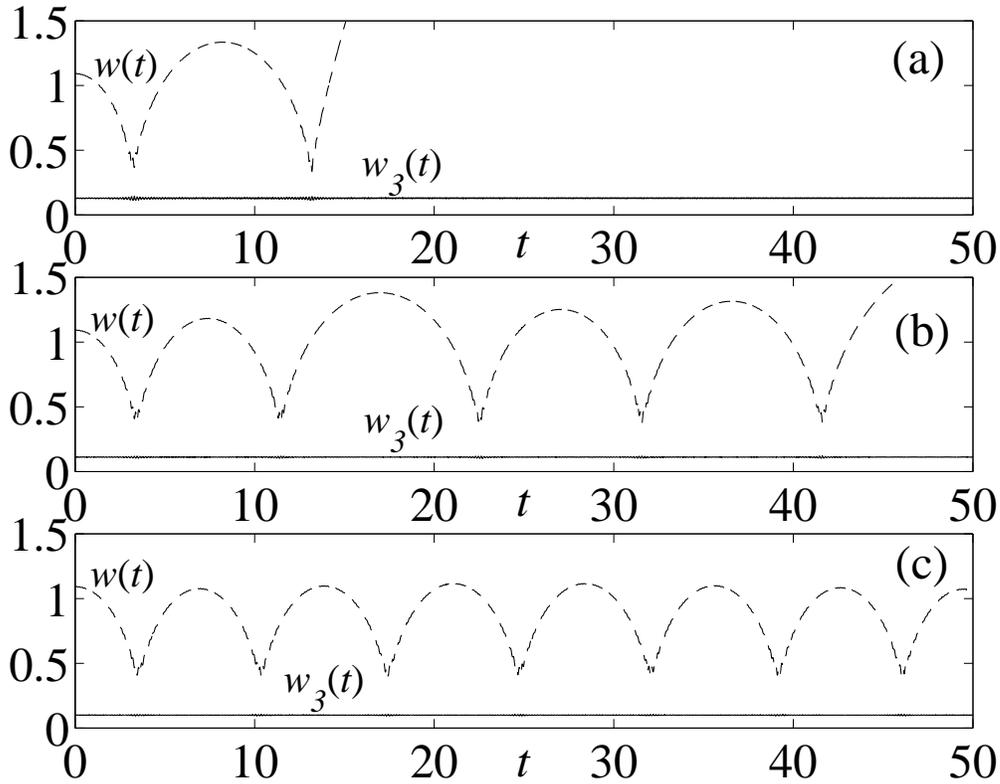,width=0.95\columnwidth}
\end{center}\caption{Stabilization of a Gaussian according to
numerical simulations of Eq. \eqref{eq:3dwidths} with parameters
$g_0=(2\pi)^{1/2} (-2.2\pi)$, $g_1=(2\pi)^{1/2} 8\pi$,
$\Omega=40$, $w(0)=1.09$, $w_3(0)=w_{3,0}=\lambda_3^{-1/2}$. The
evolution of the widths $w$ (dashed) and $w_3$ (solid) is shown
for (a) $\lambda_3=60$, (b) $\lambda_3=80$ and (c)
$\lambda_3=100$. \label{lambdas}}
\end{figure}

In Fig. \ref{lambdas} we plot the evolution of the widths after
solving numerically Eqs. \eqref{eq:3dwidths} for different values
of $\lambda_3$. We have taken $g_0=(2\pi)^{1/2} (-2.2\pi)$,
$g_1=(2\pi)^{1/2} 8\pi$, $\Omega=40$, $w(0)=1.09$ and
$w_3(0)=w_{3,0}=\lambda_3^{-1/2}$. We expect that the values of
$w$ during the evolution will be about $w=1$, as in 2D case, so
the condition \eqref{eq:lambda} says that $\lambda_3 \gg 5$,
approximately. We can see that the greater the value of
$\lambda_3$ is the better the stabilization is, according to our
previous estimates.

We have compared these results with the simulations of the full
NLSE for the case when the potential is restricted to the $z$-axis
and the system is strongly coin-shaped. To do so we have used a
pseudo-spectral fully asymmetric 3D evolution method combined with
a second order split-step method to advance in time. Typical grid
sizes were of $128\times 128\times 64$.

\begin{figure}
\begin{center}
\epsfig{file=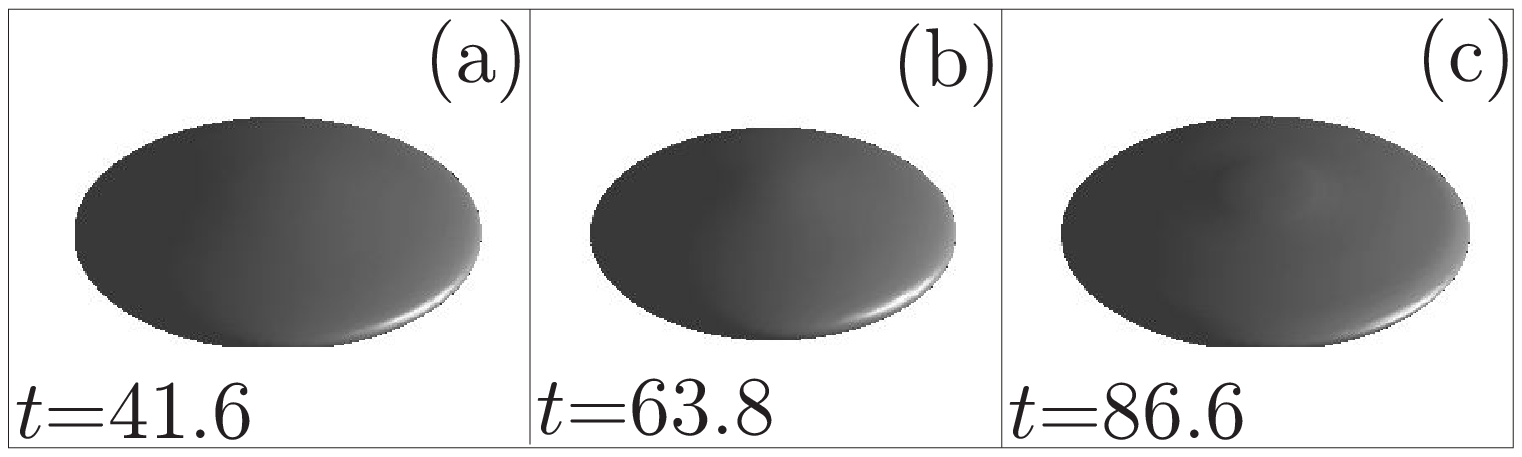,width=0.95\columnwidth}
\epsfig{file=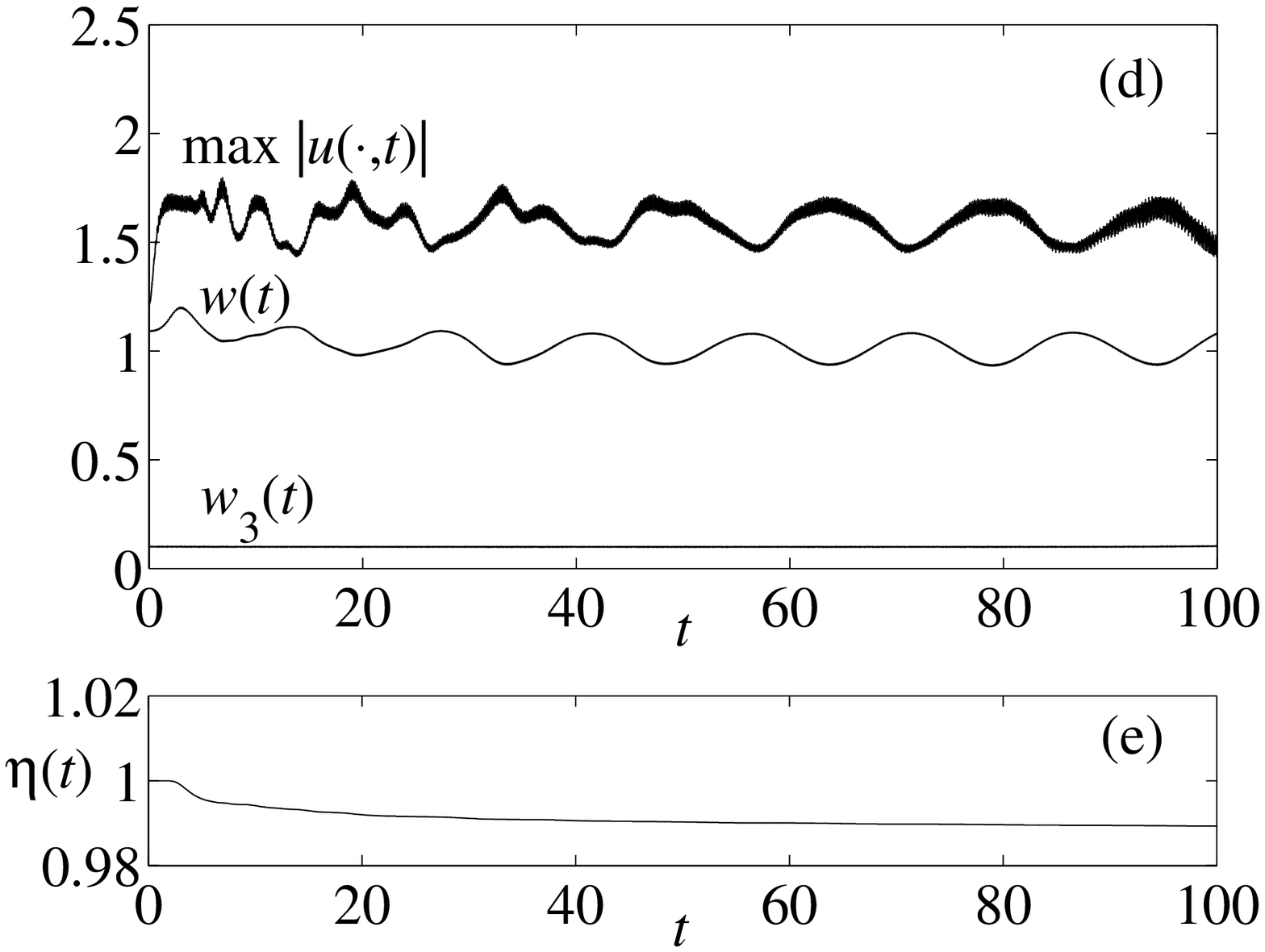,width=0.95\columnwidth}
\end{center}\caption{Results of numerical simulations of Eq.
\eqref{eq:gpe2} showing stabilization of Gaussian initial data
$\psi({\bf x},0)=(1/\pi^{3/4}w(0)w_3(0)^{1/2}) e^{-(x^2+y^2)/2
w(0)^2} e^{-z^2/2 w_3(0)^2}$ with widths $w(0)=1.09$ and
$w_3(0)=\lambda_3^{-1/2}$ for parameter values $g_0=(2\pi)^{1/2}
(-2\pi)$, $g_1=(2\pi)^{1/2} 8\pi$, $\Omega=40$ and
$\lambda_3=100$. (a-c) Isosurface plots of $|u({\bf x},t)|^2$ for
$|u|^2=0.01$ corresponding to times of maximum (a,c) and minimum
(b) width $w(t)$ of the solution. (d) Evolution of the widths
$w(t)$ and $w_3(t)$ and of the amplitude $\max |u(\cdot,t)|$ of
the solution. (e) Evolution of the norm $\eta(t)$.
\label{quasi2d}}
\end{figure}

In Fig. \ref{quasi2d} we present the results for $g_0=(2\pi)^{1/2}
(-2\pi)$, $g_1=(2\pi)^{1/2} 8\pi$, $\Omega=40$, $\lambda_3=100$
and as initial data we take a Gaussian with $w(0)=1.09$ and
$w_3(0)=\lambda_3^{-1/2}$. We see that, as in the two-dimensional
case, part of the wave packet goes outside and after a
readjustment the solution oscillates in the same way that in two
dimensions. Moreover, the width $w_3$ remains nearly constant
during the evolution.

\section{Conclusions and discussion}
\label{VI}

In this paper we have provided a deeper understanding of the
phenomenon of stabilization of solitons of the nonlinear
Schr\"{o}dinger equation by means of the control of the nonlinear
term. We have developed the moment equations for 2D and 3D systems
and obtained, on the basis of their rigorous analysis, precise
conditions for the stabilization of 2D systems.

Taking as initial data Townes solitons or Gaussians for
simulations based on Eq. \eqref{eq:gpe3} we have shown that the
former is the structure which is stabilized and that other initial
data which can be trapped must eject a fraction of the wave packet
in the form of radiation to accomodate to this specific solution.

Also we have analyzed the three-dimensional situation. Here we
have made an extensive search of stable regions according to our
moment equations improving and extending the analysis of Ref.
\cite{pisaBoris}. We have justified why moment-type equations
cannot be used to predict the dynamics of the system. Only limited
time stabilization is possible when the parameters are fine-tuned
to very precise values. Finally, when a strong trapping along one
specific direction is kept the system becomes effectively
two-dimensional and it can be described again by variational
methods whose predictions agree well with the full numerical
simulations of the problem. In the latter case three-dimensional
confinement is possible although there are only
 two spatial directions along which the solution is trapped by
 the nonlinear forces plus the stabilization mechanisms while the
 other direction is trapped by harmonic forces provided by the
 potential.

 It is remarkable that the nonlinear Schr\"{o}dinger equations
 support these stabilized structures which open new fields for
 applications, in fact these are the first stable structures
 obtained in the framework of the cubic nonlinear Schr\"odinger
 equation.

 Many extensions of this work are possible. First, it would be very interesting to study the
 robustness of these breather-type solutions under different
 perturbations, e.g. by mutual collision of different structures
 in single and multicomponent systems. Secondly it could be
 interesting to try to stabilize other stationary structures
 different from the two-dimensional Townes soliton.

 \textbf{Acknowledgments}

This work has been partially supported by the Ministerio de
Ciencia y Tecnolog\'{\i}a under grants BFM2000-0521, BFM2002-01308
and Consejer\'{\i}a de Ciencia y Tecnolog\'{\i}a de la Junta de
Comunidades de Castilla-La Mancha under grant PAC02-002. G. D.
Montesinos is supported by Ministerio de Educaci\'on, Cultura y
Deporte under grant AP2001-0535.


\end{document}